\documentclass[conference]{IEEEtran}

\ifCLASSINFOpdf
  \usepackage{graphicx}
\fi
\usepackage{subfigure}
\usepackage{stfloats}
\usepackage{epstopdf}
\usepackage{bm}
\usepackage{stfloats}
\usepackage{amssymb}
\usepackage{algorithm}
\usepackage{algorithmic}

\IEEEoverridecommandlockouts

\usepackage[cmex10]{amsmath}
\usepackage{amssymb}
\usepackage{array}
\usepackage{mathtools}

\newtheorem{corollary}{\textbf{Corollary}}

\newtheorem{theorem}{\textbf{Theorem}}

\begin{document}


\title{Guard Region Model for Pilot Reuse Analysis in Uplink Massive MIMO Systems}

\author{\IEEEauthorblockN{Yuwei Ren\IEEEauthorrefmark{1}, Yingzhe Li\IEEEauthorrefmark{2}, Jeffrey G. Andrews\IEEEauthorrefmark{2}, and Yingmin Wang\IEEEauthorrefmark{3}}

\IEEEauthorblockA{\IEEEauthorrefmark{1}School of Information and Communication Engineering, Beijing University of Posts and Telecommunications, China}

\IEEEauthorblockA{\IEEEauthorrefmark{2}Department of Electrical \& Computer Engineering, The University of Texas at Austin, USA}

\IEEEauthorblockA{\IEEEauthorrefmark{3}State Key Laboratory of Wireless Mobile Communications, China Academy of Telecommunications Technology, China}

Email: yuweir@gmail.com, yzli@utexas.edu, jandrews@ece.utexas.edu, wangyingmin@catt.cn}

\maketitle

\begin{abstract}
Massive multiple-input multiple-output (MIMO) are expected to significantly enhance the spectrum efficiency (SE) and energy efficiency (EE) of future cellular systems. Since the performance gain of massive MIMO is fundamentally limited by pilot contamination, pilot reuse design is crucial to achieve reasonable cell throughput and user rate. In this paper, we leverage stochastic geometry to model pilot reuse in massive MIMO systems by introducing a guard region model to match realistic pilot allocation strategies. The uplink signal-to-interference-plus-noise ratio (SINR) distribution is analytically derived, based on which the benefits of pilot reuse on cell-throughput and user-rate are investigated. The optimal pilot reuse factor for uplink transmission is obtained. We also find through simulations that increasing the pilot reuse factor beyond a certain value would not improve user-rate, and could even lead to a significant loss of the cell throughput.
\end{abstract}

\section{Introduction}
\begin{figure*}
\centering
\subfigure[Random Model.] { \label{Fig:a}
\includegraphics[width=0.646\columnwidth]{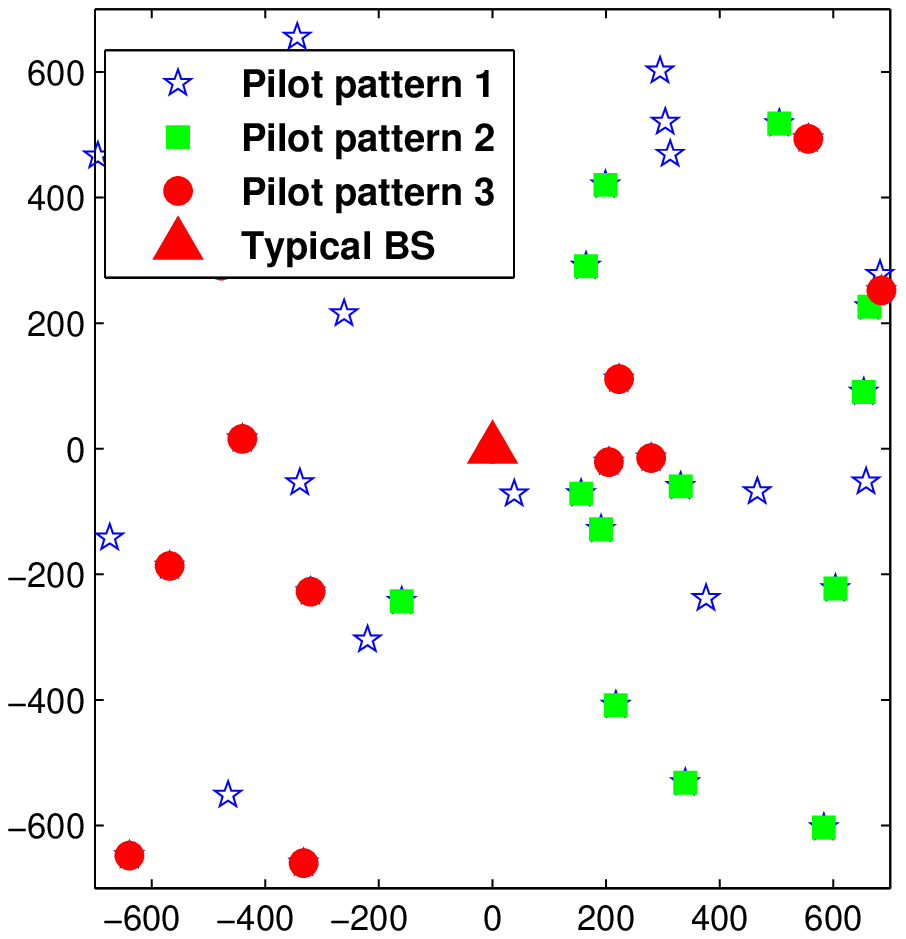}
}
\subfigure[Hexagonal Model.] { \label{Fig:b}
\includegraphics[width=0.646\columnwidth]{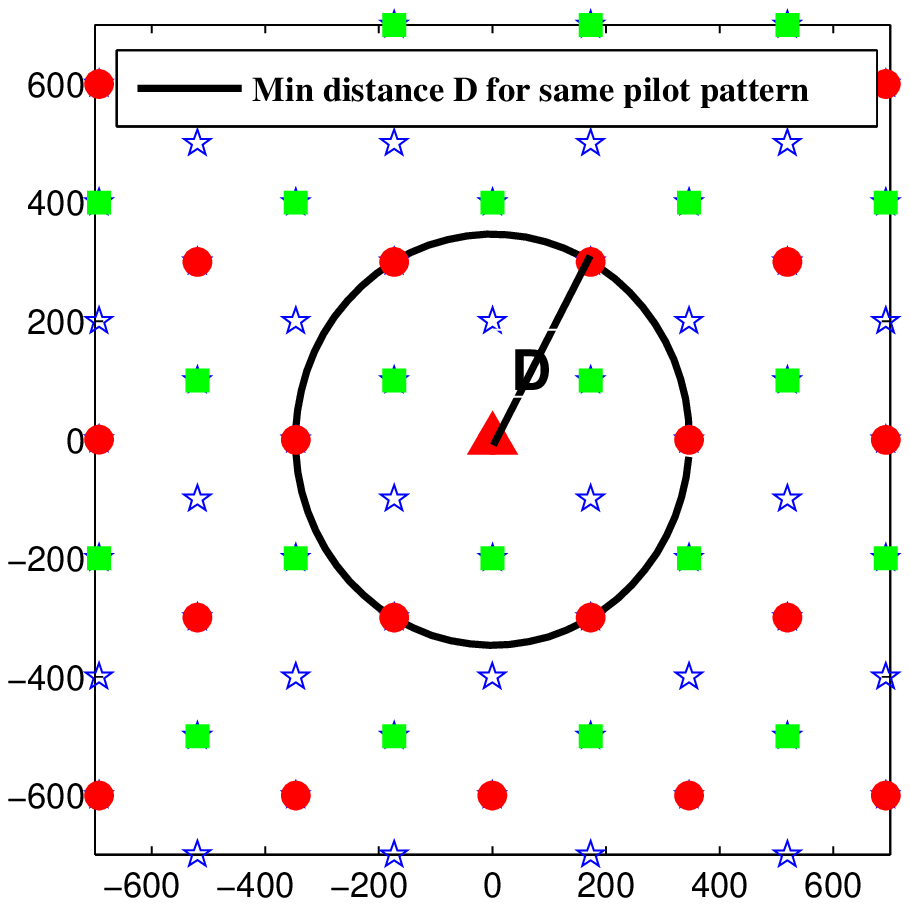}
}
\subfigure[Guard Region Model.] { \label{Fig:c}
\includegraphics[width=0.646\columnwidth]{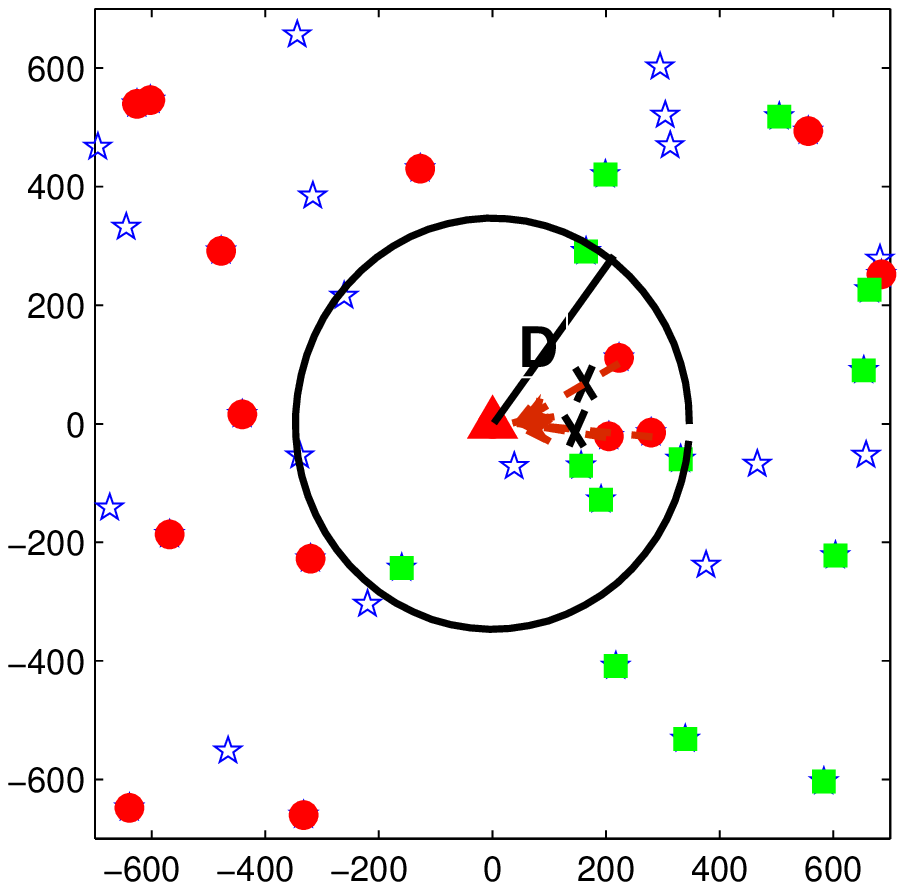}
}
\caption{Illustration of one realization of BS position from guard region PPP model, compared to random PPP model and ideal hexagonal model, with the same BS density (${\lambda_B}=2.8 \times{10^{-5}}$ [BSs per $m^2$]), $K=10$ per cell and pilot reuse factor $\Delta=3$.}
\label{fig1}
\end{figure*}

Massive MIMO system is considered as a scalable advanced architecture that would play a key role in future 5G cellular networks \cite{jeff-5G}. Some simple linear schemes could achieve high capacity performance for massive MIMO systems.
For example, maximum ratio transmission (MRT) in the downlink and maximum ratio combining (MRC) in the uplink are able to achieve  comparable performance to optimal nonlinear schemes \cite{Marzetta} when the number of base station (BS) antennas become large.

Despite massive MIMO's potential benefits, it is still subject to several practical implementation challenges.
The first challenge is cost, which includes hardware, deployment and computation (delay) costs.
For example, the one-bit analog-to-digital converter (ADCs) is applied in \cite{Jianhua1} to reduce the cost of front-end antennas. Low-complexity precoding methods are considered in \cite{Yuwei} to reduce the computation cost.

Another challenge is the pilot contamination. Specifically, pilot contamination refers to the fact that the SINR of massive MIMO systems is ultimately limited by the large-scale channel information from the BSs with common pilots \cite{Marzetta}. Several recent works have investigated pilot contamination. For example, the performance of precoding in a multi-cell TDD system with pilot contamination is studied in \cite{Jubin}, which shows that multi-cell cooperation precoding gets better performance with non-orthogonal pilot sequences.
The relationship between pilots and users is investigated in \cite{Emil-2015}. Its asymptotic analysis leads to the optimum number of users given antennas and cell structures. Several mitigating approaches to pilot contamination have been proposed in \cite{Abbs6} and its references, but they introduce significant complexity.

All the aforementioned works are based on a simplified network topology, e.g., considering only a few base stations in a ideal hexagonal grid. Since realistic deployments of cellular networks are typically irregular~\cite{li2015statistical} and extend far beyond a few neighbouring cells, it is of great interest to analyze non-regular network topologies. Fortunately, by modeling the BS deployment as a realization of Poisson point processes (PPPs), stochastic geometry is able to facilitate mathematical characterizations of the signal-to-interference ratio (SIR) distribution in cellular networks with single-antenna BSs  \cite{Jeff-2011}. The stochastic geometry analysis has been extended to massive MIMO systems in \cite{tianyang-9,tianyang-10}, but the derived asymptotic SIR is approached with impractically large number of antennas, e.g. ${10^4}$ antennas. In addition, \cite{tianyang-jour} shows the relationship between the number of users and number of antennas in an uplink masive MIMO system without considering pilot reuse.

The pilot reuse design for multi-cell massive MIMO cellular system is considered in \cite{Emil-2015} under an ideal hexagonal BS location model, as shown in Fig. \ref{Fig:b}. Due to the symmetric topology in hexagonal model, the optimal pilot reuse factor found in \cite{Emil-2015} is restricted to certain values (e.g., 1,3,4,7,...). In addition, \cite{Emil-2015Jun} gives a lower bound on the energy efficiency in a Poisson massive MIMO network as shown in Fig. \ref{Fig:a}, which assumes a sub-optimal pilot reuse approach such that the pilot allocated to each BS is i.i.d. uniformly distributed.

In this paper, we propose a novel pilot reuse model for an uplink massive MIMO system, and we apply stochastic geometry to study the pilot contamination problem with MRC.
Our main contributions can be summarized as follows:
\begin{itemize}
  \item We introduce a ``guard region" \cite{Robert} to analyze pilot contamination in uplink, abiding by the basic Physical Cell Identifier (PCI) planning rules \cite{PCIplanning} to ensure the minimum pilot contamination distance.
  \item The proposed model changes the distribution of users, which is amenable to stochastic geometry analysis. We derive the analytical uplink SINR distribution of a massive MIMO system with a finite number of antennas.
  \item Based on the analytical and simulation results, we give the relationship between pilot reuse, SIR distribution and cell-throughput.
\end{itemize}

\section{System Model}
We consider a cellular network that is designed to serve a uniform user distribution, where the BSs are distributed in ${\mathbb{R}^2}$ according to a homogeneous PPP ${\Phi}$ with intensity ${\lambda_B}$.
Each BS has $M$ antennas and serves $K$ single-antenna user equipments (UEs).
Each UE connects to its closest BS, hence the coverage area of a BS is its Poisson-Voronoi cell.
Assume the pilot reuse factor (i.e., number of pilot groups) is $\Delta$, then there are $\Delta \times K$ orthogonal pilot sequences to support uplink channel training in TDD model. Each cell selects one set of pilots from $\Delta$ groups, which means that users share different pilots in the same cell, and pilot contamination might exist among different cells.

Considering the drawbacks of the random model in Fig. \ref{Fig:a} and ideal hexagonal model in Fig. \ref{Fig:b}, this section first introduces a more tractable model to evaluate pilot contamination, namely the “guard region model”.
Next, the corresponding channel model, power control model and uplink transmission model are presented.

\subsection{Guard Region Model}
Most existing works use the ideal hexagonal cellular model for pilot contamination, where the pilot resource is allocated by some symmetrical methods. Fig. \ref{Fig:b} shows an ideal hexagonal model with pilot reuse factor $\Delta  = 3$, which keeps the minimum distance among BSs sharing the same pilot resource as $2R\sqrt \Delta$ with $R$ being the cell radius.
Although it could easily model the pilot allocation under PCI planning rules \cite{PCIplanning}, this model suffers from being both highly idealized and not very tractable.

The stochastic geometry model in Fig. \ref{Fig:a} is widely accepted as a reasonable approximation to realistic deployments, and is tractable to analyze key system performance metrics. Due to the irregular deployment of BSs, the commercial PCI planning needs to utilize exhaustive system simulation to search a reasonable solution, which must follow the \emph{non-collision} and \emph{non-confusion} rules \cite{PCIplanning}. However, most existing works on stochastic geometry analysis of massive MIMO ignore these basic rules.

In order to abide by the PCI planning rules, we define the guard region as a circular area with radius $D$ around the typical BS, which is shown in Fig. \ref{Fig:c}. There is no pilot contamination to the BS inside the guard region, while it suffers pilot contamination with random probability $\frac{1}{\Delta}$ outside the guard region. Compared to the complete random model in Fig. \ref{Fig:a}, the typical BS in Fig. \ref{Fig:c} is not subject to pilot contamination from its three nearest neighboring BSs, which is consistent with the commercial PCI planning. In addition, for the ideal hexagonal model in Fig. \ref{Fig:b}, the pilot allocation scheme and minimum distance $D$ are always fixed and optimal, which could be treated as the upper bound to the realistic model.
In contrast, the guard region model simultaneously complies with the random distribution of BSs and the basic PCI planning rules. Moreover, the guard region radius $D$ can be adjusted according to the SIR constraint per UE and/or PCI planning methods.
In order to illustrate the difference for these three models, we give the simulation results in Fig. \ref{fig2} in Section VI.

\subsection{ Uplink Channel and Power Control Model}
\setcounter{equation}{4}
\begin{figure*}
\centering
	\begin{equation}\label{RxSigEq}
    \scalebox{0.9}{$
	\widehat {\bf{h}}_{00k}^H{{\bf{y}}_{00k}} = {x_{00k}}{\rm E}\left\{ {\widehat {\bf{h}}_{00k}^H{{\bf{h}}_{00k}}} \right\} + \underbrace {{x_{00k}}\left( {\widehat {\bf{h}}_{00k}^H{{\bf{h}}_{00k}} - {\rm E}\left\{ {\widehat {\bf{h}}_{00k}^H{{\bf{h}}_{00k}}} \right\}} \right) + \sum\limits_{j \ne k}^K {\widehat {\bf{h}}_{00k}^H{{\bf{h}}_{00j}}{x_{00j}}}+ \sum\limits_{i \in {\Psi _\lambda }}^{} {\sum\limits_{j = 1}^K {\widehat {\bf{h}}_{00k}^H{{\bf{h}}_{0ij}}{x_{0ij}}} }  + \widehat {\bf{h}}_{00k}^H{\bf{n}}}_{{{\bf{z}}_{00k}}},
    $}
	\end{equation}
\end{figure*}
\begin{figure*}
\centering
	\begin{equation}\label{SINR-1}
    \scalebox{0.9}{$
	\begin{array}{l}
	SINR \approx \frac{{{p_{00k}}{{\left| {{\rm E}\left\{ {\widehat {\bf{h}}_{00k}^H{{\bf{h}}_{00k}}} \right\}} \right|}^2}}}{{{\rm E}\left\{ {{{\left| {{{\bf{z}}_{00k}}} \right|}^2}} \right\}}}  = \frac{{{p_{00k}}{{\left| {{\rm E}\left\{ {\widehat {\bf{h}}_{00k}^H{{\bf{h}}_{00k}}} \right\}} \right|}^2}}}{{{p_{00k}}Var\left\{ {\widehat {\bf{h}}_{00k}^H{{\bf{h}}_{00k}}} \right\} + \sum\limits_{j \ne k}^K {{p_{00j}}{\rm E}\left\{ {{{\left| {\widehat {\bf{h}}_{00k}^H{{\bf{h}}_{00j}}} \right|}^2}} \right\}}  + \sum\limits_{i \in {\Psi _\lambda }}^{} {\sum\limits_{j = 1}^K {{p_{0ij}}{\rm E}\left\{ {{{\left| {\widehat {\bf{h}}_{00k}^H{{\bf{h}}_{0ij}}} \right|}^2}} \right\}} }  + {\sigma ^2}{\rm E}\left\{ {{{\left| {\widehat {\bf{h}}_{00k}^H} \right|}^2}} \right\}}},
	\end{array}
    $}
	\end{equation}
\hrulefill
\end{figure*}

\setcounter{equation}{0}
We consider a TDD system with perfect synchronization, and we denote by ${{\rm T}_{g,k}}$ the pilot sequence for $k$-{th} user in the $g$-{th} pilot resource group, where $k\in\left[{1,K}\right]$ and $g\in\left[{1,\Delta}\right]$. In the uplink channel training stage, the scheduled users send their assigned pilots ${{\rm T}_{g,k}}$, and base station estimates the channels using the corresponding orthogonal pilots. In the uplink transmission phase, the BSs apply MRC to receive the uplink data, based on the estimated channel.

The channel is assumed constant during one resource block and fades independently from block to block. This narrow-band channel model which could be guaranteed by applying orthogonal frequency-division multiplexing (OFDM) and frequency domain equalization \cite{GoldSmith-book}.
Here, we denote the channel model by:
\begin{equation}\label{Channel-model}
  {{\bf{h}}_{lnk}}={\left({{\beta_{lnk}}}\right)^{1/2}}{{\mathbf{w}}_{lnk}},
\end{equation}
where ${\beta_{lnk}}$ is the large-scale path loss for the $l$-{th} BS to the $k$-{th} UE associated with the $n$-{th} BS, and ${{\mathbf{w}}_{lnk}}$ is a Gaussian vector with the distribution $CN\left( {{\mathbf{0,}}{{\mathbf{I}}_M}} \right)$ (i.e., complex i.i.d. Rayleigh fading).

The large-scale path loss gain ${\beta_{lnk}}$ is computed as
\begin{equation}\label{large scale}
  {\beta_{lnk}}=C{\left({\max\left({{R_{lnk}},\delta}\right)}\right)^{-\alpha}},
\end{equation}
where $C$ is a constant determined by the carrier frequency and reference distance, ${R_{lnk}}$ is the corresponding distance, $\alpha>2$ is the path loss exponent, and $\delta\geqslant 0$ is the reference distance (e.g. 1 meter), intended to address the near field effect.
Similar path loss models have been used in prior work on cellular network analysis \cite{Jeff-2011}.

We utilize the fractional power control as in LTE \cite{Tianyang-13} in both the uplink training and uplink data stages. Specifically, the user ${y_k}$ of the $n$-th BS transmits with power ${P_{nnk}}={P_t}{\left( {{\beta_{nnk}}}\right)^{-\varepsilon }}$, where $\varepsilon\in\left[{0,1}\right]$ is the fraction of the path loss compensation, and ${P_t}$ is the open loop transmit power with no power control.
Further, we omit the constraint on the maximum uplink transmit power and set ${P_t}=1$ for simplicity.
To maintain tractability, we assume that the BSs estimate the channel by correlating the received training signal with the corresponding pilot without using minimum mean squared error estimation.

Hence, the estimated channel in the channel estimation stage is given by:
\begin{equation}\label{estimated channel}
  {\widehat {\mathbf{h}}_{00k}} = \sqrt {{P_{00k}}} {{\mathbf{h}}_{00k}} + \sum\limits_{i \in {\Psi _\lambda }}^{} {{\zeta_{ik}}\sqrt {{P_{iik}}} {{\mathbf{h}}_{0ik}}},
\end{equation}
where $\sum\limits_{i\in{\Psi_\lambda}}^{}{{\zeta_{ik}}\sqrt{{P_{iik}}}{{\mathbf{h}}_{0ik}}}$ is the estimation error caused by pilot contamination, ${\Psi_\lambda }$ is the set of BSs excluding the typical BS, and ${\zeta_{ik}}\in\left\{{0,1}\right\}$ is a Bernoulli random variable with parameter $\frac{1}{\Delta}$.

Therefore, the uplink received signal after MRC is:
\begin{equation}\label{Uplink received signal}
\scalebox{0.83}{
$\begin{array}{l}
\widehat {\bf{h}}_{00k}^H{{\bf{y}}_{00k}}\\
 = \widehat{\bf{h}}_{00k}^H\left({{{\bf{h}}_{00k}}{x_{00k}}+\!\!\!\!\!\!\sum\limits_{j = 1, j \ne k}^K {{{\!\!\!\bf{h}}_{00j}}{x_{00j}}}+\!\!\!\sum\limits_{\varphi\in {\Psi_\lambda}}\!{\sum\limits_{j=1}^K {{{\bf{h}}_{0\varphi j}}{x_{\varphi \varphi j}}} }+{\bf{n}}}\right),
\end{array}$
}
\end{equation}
where ${x_{\varphi\varphi j}}$ is the transmitted signal with distribution $CN\left( {0,{P_{\varphi \varphi j}}} \right)$, and $\bf{n}$ is the noise with distribution $CN\left( {0,{\sigma ^2}{{\bf{I}}_M}} \right)$.
Eq. (\ref{Uplink received signal}) represents a simple massive MIMO system in the uplink with MRC, in which the estimation normalization could be ignored, and the $SINR$ expression can be analytically derived using stochastic geometry.

\section{Performance Analysis with MRC Receivers}
In this section, we derive the approximate SINR distribution, investigate the effect of pilot reuse factor on SINR performance, and obtain the cell throughput.
\setcounter{equation}{6}
\subsection{Approximate SINR Analysis}
The pilot contamination and interference in (\ref{Uplink received signal}) make the SINR distribution intractable. Therefore, we derive an approximate SINR distribution by simplifying the numerator and denominator similar to Theorem 1 in \cite{Jubin}. Specifically, we rewrite the received uplink signal given by (\ref{Uplink received signal}) into (\ref{RxSigEq}), and define ${x_{00k}}{\rm E}\left\{{\widehat{\bf{h}}_{00k}^H{{\bf{h}}_{00k}}}\right\}$ as the desired signal, and ${{\bf{z}}_{00k}}$ as the additive noise.
The desired signal only depends on the channel distribution, rather than the instantaneous channel.
The additive noise is neither independent nor Gaussian, and we apply the approximation theorem in \cite{Jubin-34} to model the worst-case uncorrelated additive noise, such that the desired signal ${x_{00k}}{\rm E}\left\{ {\widehat {\bf{h}}_{00k}^H{{\bf{h}}_{00k}}} \right\}$ is dependent to the noise ${{\bf{z}}_{00k}}$. As a result, an approximate SINR can be expressed as (\ref{SINR-1}), whose distribution is given by:
\begin{theorem}\label{SINRThm}
With i.i.d. Rayleigh fading and fractional power control, the uplink $SINR$ distribution can be approximated by:
\begin{equation}\label{SINR-Prob}
\scalebox{0.88}{$
\begin{array}{l}
\mathbb{P}\left[{SINR > T}\right]\approx\\
\sum\limits_{n=1}^N {\left({\begin{array}{*{20}{c}}
N\\n
\end{array}}\right){{\left({-1}\right)}^{n+1}}{e^{-\eta Tn\frac{1}{M}}}}\int_0^\infty{Z\left(x\right)f\left(x\right){{\left[{L\left(x\right)}\right]}^{K-1}}}{\rm d}x,
\end{array}
$}
\end{equation}
where we have:
\begin{align*}
&f\left( x \right) = 2\pi x{\lambda _B}{e^{ - \pi {\lambda _B}{x^2}}},\\
&Z\left( x \right) = {e^{ - \eta Tn\frac{1}{M}\left( {A{x^{\alpha \left( {1 - \varepsilon } \right)}} + B{x^{2\alpha \left( {1 - \varepsilon } \right)}}} \right)}},\\
&L\left( x \right) = \int_0^\infty \!\!  {\left[ {\frac{e^{{C_1}\left( x \right){u^{ - \frac{\alpha }{2}(1 - \varepsilon )}}}}{\Delta } + \left( {1 - \frac{1}{\Delta }} \right){e^{{C_2}\left( x \right){u^{ - \frac{\alpha }{2}(1 - \varepsilon )}}}}} \right]{e^{ - u}}} {\rm d}u,
\end{align*}
and the following notations:
\begin{align*}
&A = \frac{{{P_1}{I_1}}}{\Delta } + \left( {K} \right){P_1}{I_1} + {\sigma ^2},\\
&B = \frac{{\left( {M + 1} \right){P_2}{I_2} + {P_1}{I_1}A - {{\left( {{P_1}{I_1}} \right)}^2}}}{\Delta },\\
&{C_1}\left( x \right) = {\mathbb{N}}\left( {{x^{\alpha \left( {1 - \varepsilon } \right)}} + {P_1}{I_1}{x^{2\alpha \left( {1 - \varepsilon } \right)}}} \right),\;{C_2}\left(x\right) = {\mathbb{N}}{x^{\alpha \left( {1 - \varepsilon } \right)}}.
\end{align*}
In addition, $N$ is the number of terms used in the approximation,
$\eta= N{\left( {N!} \right)^{ - \frac{1}{N}}}$,
${\Bbb N}=-\eta Tn\frac{1}{M}{\left({\pi{\lambda _B}}\right)^{\frac{\alpha }{2}\left({1-\varepsilon}\right)}}$.
We also have ${P_\omega } = {\left( {{\lambda _b}\pi } \right)^{ - \alpha \varepsilon \omega /2}}\Gamma \left( {\frac{{\alpha \varepsilon \omega }}{2} + 1} \right)$, ${{I}_{\omega }}=\frac{{{\lambda }_{b}}\pi }{1-\alpha {{2}^{\omega -2}}}{{D}^{2-\omega \alpha }}$ for $\omega  \in \left\{ {1,2} \right\}$, where $\Gamma \left(  \bullet  \right)$ is the Gamma function.
\end{theorem}

\emph{Proof:}
First, get the expectations respectively in numerator and denominator.
And by expectation, the SINR is displaced only by large-scale information,
and $\mathbb{P}\left[{SINR > T}\right]$ can be calculated as in Appendix A.
Meanwhile, next section certifies the accuracy of the approximation in Theorem 1.

\subsection{Effect of Pilot Reuse on SINR Performance}
In this part, based on the approximate SINR distribution, we investigate how pilot reuse factor affects the SINR performance under the guard region model. We assume the system is interference limited, i.e., ${{\sigma }^{2}}=0$.
In order to compare with the optimal hexagonal structure, we define the minimum distance $D=2R\sqrt{\Delta }$, which relates the guard region size to the pilot reuse factor $\Delta$.

Since the interference is mostly contributed by the nearest BSs using the same pilot resources, a larger pilot reuse factor $\Delta$ would lead to smaller interference. Meanwhile, a larger $\Delta$ will also incur higher overhead for pilot resources. Therefore, one rule for selecting $\Delta$ is to ensure that the probability for SINR to be greater than a specified threshold $T$ is sufficiently large (e.g., larger than some value $\gamma$), which can be expressed as follows:
\begin{equation}\label{Corollary-1}
  \mathbb{P}\left[{SINR > T}\right]\ge\gamma.
\end{equation}
Here $T$ represents a minimum SINR threshold to ensure certain basic rate for users. Based on (\ref{Corollary-1}), we can obtain the following relation between the SINR requirement $\gamma$ and pilot reuse factor $\Delta$:

\begin{corollary}~\label{PilotFactorCoro}
Under guard region model with $D=2R\sqrt{\Delta }$, the minimum number pilot reuse factor $\Delta$ required to satisfy~(\ref{Corollary-1}) is given by:
\begin{equation}\label{Corollary-1-1}
\Delta = {{y}^{-1}}\left( \gamma ,T \right),
\end{equation}
where
\begin{equation}\label{Corollary-2}
\begin{array}{l}
y(\Delta ) = \\
\sum\limits_{n = 1}^N {\left(\!\!\! {\begin{array}{*{20}{c}}
N\\
n
\end{array}} \!\!\! \right){{\left( { - 1} \right)}^{n + 1}}{e^{ - \eta Tn\frac{1}{M}}}} \int_0^\infty  {{{\left[ {{Q_1}(t,\Delta )} \right]}^{K - 1}}{e^{ - {Q_2}(t,\Delta ) - t}}} dt
\end{array}
\end{equation}
and
${Q_1}(t,\Delta ) = L\left( {\sqrt {\frac{t}{{\pi {\lambda _B}}}} } \right)$, $L\left(  \bullet  \right)$ is defined in Theorem 1; And
\begin{equation}\label{Corollary-2}
\begin{array}{l}
{Q_2}(t,\Delta ) = \frac{{ - \eta Tn}}{M}\left( {K + \frac{1}{\Delta }} \right){B_1}{t^{\frac{\alpha }{2}\left( {1 - \varepsilon } \right)}} + \\
\frac{{ - \eta Tn}}{{M\Delta }}\left( {\left( {M + 1} \right){B_2} + \left( {K} \right){{\left( {{B_1}} \right)}^2}} \right){t^{\alpha \left( {1 - \varepsilon } \right)}}
\end{array}
\end{equation}
Here, we define
${B_\omega } = \Gamma \left( {\frac{{\omega \alpha \varepsilon }}{2} + 1} \right)\frac{{{{\left( {\frac{{\sqrt 3 }}{{2\Delta \pi }}} \right)}^{{2^{\omega  - 1}}\alpha  - 1}}}}{{{2^{\omega  - 1}}\alpha  - 1}}$.
\end{corollary}
\emph{Proof:}
The process is similar to the Theorem 1, where changing the variable as $t=\pi {\lambda_B}{x^2}$.

From Corollary~\ref{PilotFactorCoro}, the performance is not related the density of BS. But the distribution is constrained by the number of users per cell $K$ and the pilot reuse factor $\Delta$. Next, we would like to analyze $K$ and $\Delta$ and see how they affect the performance.

\subsection{Rate Analysis}
In this section, we apply the approximate SINR results to compute the achievable rate. First, we define the average user-achieved spectrum efficiency as:
\begin{equation}\label{rate}
  {{\tau }_{0}}=E\left[ \log \left( 1+\min \left\{ SIR,{{T}_{\max }} \right\} \right) \right],
\end{equation}
where ${{T}_{\max }}$ is a SINR distortion threshold. Denote SIR coverage probability in Theorem~\ref{SINRThm} by ${P}_{C}(T)$, the average user-achieved spectrum efficiency can be computed as:
\begin{equation}\label{rate-2}
  {{\tau }_{0}}=\frac{1}{\ln 2}\int_{0}^{{{T}_{\max }}}{\frac{{{P}_{C}}(T)}{1+T}{{d}_{T}}}.
\end{equation}

In addition, we define the average cell throughput ${{\tau}_{S}}$ as:
\begin{align}\label{CellTputDefn}
{{\tau }_{S}}=K\left( 1-\frac{K\Delta }{{{T}_{C}}}\right){{\tau }_{0}},
\end{align}
where $\frac{K\Delta }{{{T}_{C}}}$ is the fraction of overhead, and ${{T}_{C}}$ is the length of the channel coherent time in terms of the number of symbol time. For simplicity, we only consider the overhead due to uplink channel training. We will investigate the effect of $M$, $K$ and $\Delta$ on the average cell throughput in next section.

\section{Numerical Results}
In this section, we present some numerical results based on Section II and Section III with ${\lambda_B}=2.8\times{10^{-5}}$ [BSs per $m^2$].
Firstly, we use simulation to compare the accuracy of guard region model to hexagonal and random model.
Next, we justify our analytical SINR distribution by comparing to the simulation results.
We will also investigate the effects of pilot reuse factor, number of users and antennas, on the SINR and cell throughput performance.

\begin{figure}
  \centering
  \includegraphics[width=3in]{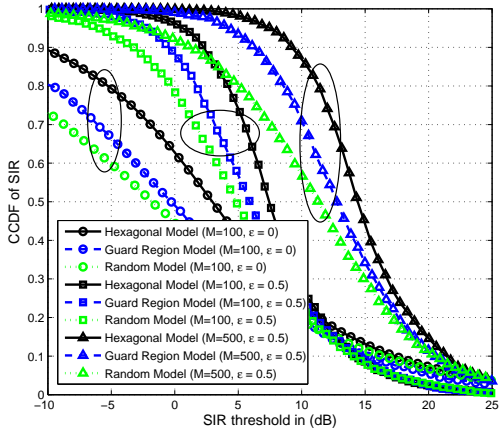}
  \caption{SIR distribution simulation for three different models with $\Delta=3$, ${\sigma ^2} = 0$, $K=10$ for $\varepsilon\in\left\{{0,0.5}\right\}$, and $M\in\left\{{100, 500} \right\}$.}\label{fig2}
\end{figure}

The SINR distribution of hexagonal, guard region, and random model under different power compensating factors $\epsilon$ and antenna numbers $M$ are compared in Fig. \ref{fig2} by simulation. In particular, we focus on the same deployment scenario as Fig. \ref{fig1},  with $K=10$ and $\Delta=3$.
We also set $D = 2R\sqrt\Delta$ to match the minimum distance in hexagonal model, and assume the noise ${\sigma ^2} = 0$.
Given $M$ and $\epsilon$, the SIR performance of the guard region model always lies within the other two models, with the hexagonal model being upper bound because of the symmetrical structure with optimal pilot reuse, and the random model being the lower bound due to lack of pilot reuse optimization. Therefore, the guard region model is a reasonable model for realistic deployment which is able to follow the PCI planning rules.

\begin{figure}
  \centering
  \includegraphics[width=3in]{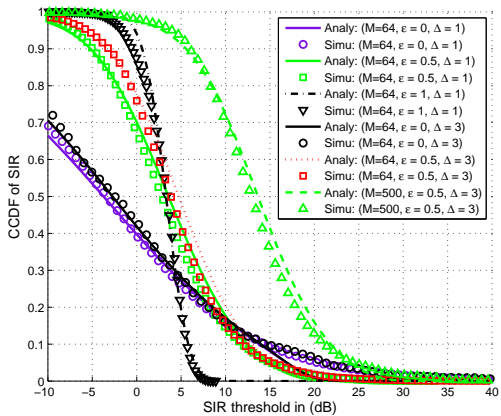}
  \caption{Comparison of theoretical and simulated SINR distribution, with ${\sigma ^2} = 0$, $K=10$ for $\varepsilon\in\left\{{0,0.5,1}\right\}$, $\Delta\in\left\{{1,3}\right\}$ and $M\in\left\{{64,500} \right\}$.}\label{fig3}
\end{figure}

The approximate SINR distribution derived in Theorem~\ref{SINRThm} and the simulation results are compared in Fig. \ref{fig3}. The accuracy of the approximations in Theorem~\ref{SINRThm} can be validated under different antenna numbers $M$, pilot reuse factor $\Delta$, and power compensation factor $\epsilon$. It can also be observed that more antennas and higher pilot reuse factor could greatly improve the SINR performance. Since the distribution of SINR becomes more centralized (e.g., closer to $5$ dB) by increasing the power compensation factor, better fairness among users can be guaranteed. In addition, even with the simple MRC in the uplink, significant SINR gains can be achieved by adding more antennas at the BS.

\begin{figure}
  \centering
  \includegraphics[width=2.9in]{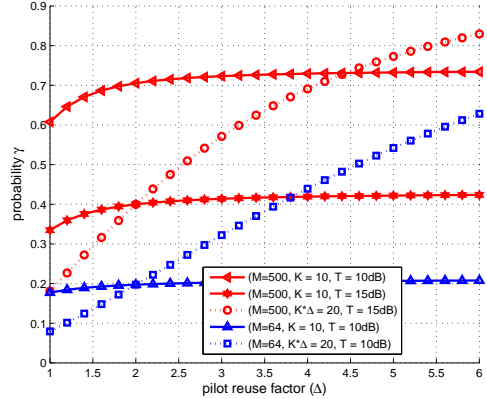}
  \caption{Relation between pilot reuse factor $\Delta$ and SINR requirement $\gamma$ in~(\ref{Corollary-1}), with ${\sigma ^2} = 0$, $\varepsilon=0.5$ for $T\in\left\{{10 \text{ dB},15 \text{ dB}}\right\}$ and $M\in\left\{{64,500} \right\}$.}\label{fig4}
\end{figure}

Fig. \ref{fig4} plots the relation between the minimum pilot reuse factor $\Delta$ and the corresponding SINR requirement $\gamma$, which is derived in (\ref{Corollary-1-1}). The effect of different antenna and user numbers, as well as the SINR threshold $T$ are considered. We can observe from Fig. \ref{fig4} that when $\Delta$ is close to 2, most cases have already reached the maximum value of $\gamma$, which means the additional pilot resources will not greatly improve the SINR.
If we set the total amout of pilot resources (i.e., $K \times \Delta$) as a constant, and decrease the number of scheduled users $K$, the probability would almost linearly increase with $\Delta$. This result shows the intra-cell interference is very crucial for SINR performance when MRC is adopted in the uplink.

\begin{figure}
  \centering
  \includegraphics[width=2.9in]{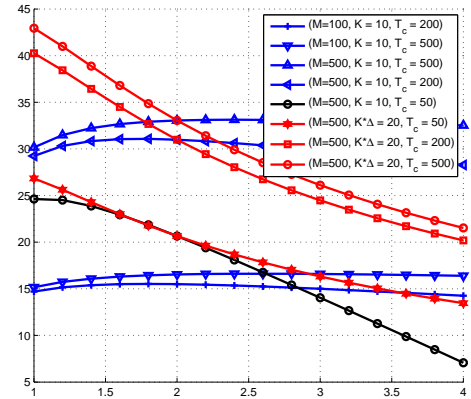}
  \caption{Average cell throughput with respect to different pilot reuse factor $\Delta$, with ${\sigma ^2} = 0$, $T_{max}=21$ dB, $\varepsilon=0.5$ for $T_{C}\in\left\{{50,200,500}\right\}$ and $M\in\left\{{100,500} \right\}$.}\label{fig5}
\end{figure}

The effect of pilot reuse factor $\Delta$ on the average cell throughput defined in~(\ref{CellTputDefn}) is investigated in Fig. \ref{fig5}. Similar to the trends in Fig. \ref{fig4}, the optimal pilot reuse factor in terms of cell throughput is typically a small number (e.g., 1, 2 or 3). When the pilot reuse factor becomes larger than that range, the cell throughput will start to decrease due to the training overhead.

As a result, the SINR and throughput performance will not benefit too much by adopting a large pilot reuse factor. In contrast, the interference from other users and antenna numbers are more decisive factors for the uplink performance of a massive MIMO system with MRC.

\section{Conclusion}
This paper studies the pilot reuse design in an uplink massive MIMO system with randomly deployed base stations. A guard region model is proposed to match the minimum distance between base stations in the hexagonal deployment that are subject to pilot contamination.
The uplink SINR distribution, effects of pilot reuse factor on SINR and cell throughput are analytically derived and numerically evaluated.
There are several potential future works, such as the analysis for downlink massive MIMO, or considering more complicated precoding and/or estimation methods.

\section{Appendix A}
First, we consider the following approximations to simply the expressions.
\begin{itemize}
\item
For Massive MIMO, based on i.i.d. matrix ${{\bf{u}}_{1 \times M}}$, these expectations could be approximated to \\
\scalebox{0.8}{
$\begin{array}{l}
E\left\{ {{{\left|{{\bf{u}}{{\bf{u}}^*}} \right|}^2}} \right\} = {M^2} + M, E{\left\{ {{{\left| {{{\bf{u}}_i}{\bf{u}}_j^ * } \right|}^2}} \right\}_{i \ne j}} = M, E\left\{ {\left| {{\bf{u}}{{\bf{u}}^ * }} \right|} \right\} = M
\end{array}$.}
\item
Assume that ${R_{llk}}$ is a Rayleigh random variable with mean $0.5\sqrt {{1 \mathord{\left/ {\vphantom {1 {{\lambda _B}}}} \right. \kern-\nulldelimiterspace} {{\lambda _B}}}}$ in \cite{Jeff-2011}.
Meanwhile, the scheduled users in other cells are modeled by \emph{exclusion ball model} \cite{Tianyang}, which the parameter $D$ ensures that they form a homogeneous PPP with density ${{\lambda _B}}$ outside the ball with radius $D$.
\end{itemize}

So we can get:
\begin{equation}\label{expectation-2}
\scalebox{0.8}{
$\begin{array}{l}
{\left( {R_{iik}^{ - \alpha }} \right)^{ - \varepsilon \omega }} \approx E\left[ {{R_{iik}}^{\omega \alpha \varepsilon }} \right]\mathop  = \limits {P_\omega }\\
\sum\limits_{i \in {\Psi _\lambda }}^{} {{{\left( {{\beta _{0ik}}} \right)}^\omega }}  \approx E\left\{ {\sum\limits_{i \in {\Psi _\lambda }}^{} {{{\left( {{\beta _{0ik}}} \right)}^\omega}}} \right\} = 2\pi {\lambda _B}\int_D^\infty  {{x^{ - \omega \alpha }}xdx}  = {I_\omega }.
\end{array}$}
\end{equation}

The numerator in (\ref{SINR-1}) is approximated as ${\left( {{\beta _{00k}}} \right)^{2 - 2\varepsilon }}{M^2}$.
The first term in the denominator is
\begin{equation}\label{expectation-3}
\scalebox{0.8}{
$\begin{array}{l}
{\left( {{\beta _{00k}}} \right)^{2 - 2\varepsilon }}M + {\left( {{\beta _{00k}}} \right)^{1 - \varepsilon }}\sum\limits_{i \in {\Psi _\lambda }}^{} {\alpha _{ik}^2{{\left( {{\beta _{iik}}} \right)}^{ - \varepsilon }}{{\left( {{\beta _{0ik}}} \right)}^1}M} \\
\mathop  \approx \limits^{(a)} {\left( {{\beta _{00k}}} \right)^{2 - 2\varepsilon }}M + {\left( {{\beta _{00k}}} \right)^{1 - \varepsilon }}\frac{M}{\Delta }E\left[ {{{\left( {{\beta _{iik}}} \right)}^{ - \varepsilon }}} \right]E\left[ {\sum\limits_{i \in {\Psi _\lambda }}^{} {{{\left( {{\beta _{0ik}}} \right)}^1}} } \right]\\
\mathop  = \limits^{(b)} M\left[ {{{\left( {{\beta _{00k}}} \right)}^{2 - 2\varepsilon }}M + {{\left( {{\beta _{00k}}} \right)}^{1 - \varepsilon }}\frac{{M{P_1}{I_1}}}{\Delta }} \right],
\end{array}$}
\end{equation}
where ${(a)}$ is from $E\left\{ {{\alpha _{ik}}} \right\} = E\left\{ {\alpha _{ik}^2} \right\} = \frac{1}{\Delta }$, ${(b)}$ is based on the definition in (\ref{expectation-2}).

Similarly, the second term is \scalebox{0.8}{$\sum\limits_{j \ne k}^K {\left( {{{\left( {{\beta _{00k}}} \right)}^{1 - \varepsilon }} + \frac{{{P_1}{I_1}}}{\Delta }} \right)} {\left( {{\beta _{00j}}} \right)^{1 - \varepsilon }}M$},
and the fourth term \scalebox{0.8}{$M{\sigma ^2}\left[ {{{\left( {{\beta _{00k}}} \right)}^{1 - \varepsilon }} + \frac{{{P_1}{I_1}}}{\Delta }} \right]$}.
The third term is
\begin{equation}\label{expectation-4}
\scalebox{0.8}{
$\begin{array}{l}
M{\left( {{\beta _{00k}}} \right)^{1 - \varepsilon }}K{P_1}{I_1} + \frac{M}{\Delta }K{\left( {{P_1}{I_1}} \right)^2} + \frac{{\left( {{M^2} + M} \right)}}{\Delta }E\left\{ {\sum\limits_{\varphi  \in {\Psi _\lambda }}^{} {{{\left( {{\beta _{\varphi \varphi k}}} \right)}^{-2\varepsilon }}{{\left( {{\beta _{0\varphi k}}} \right)}^2}} } \right\}\\
\mathop  = \limits^{(c)} M{\left( {{\beta _{00k}}} \right)^{1 - \varepsilon }}K{P_1}{I_1} + \frac{M}{\Delta }K{\left( {{P_1}{I_1}} \right)^2} + \frac{{\left( {{M^2} + M} \right){P_2}{I_2}}}{\Delta },
\end{array}$
}
\end{equation}
where ${(c)}$ is similar to ${(b)}$ with $\omega = 2$.

Next, conditioning on ${R_{00k}} = x$, we take the above expressions into (\ref{SINR-1}), and compute the conditional uplink SINR distribution as
\begin{equation}\label{ggd}
\scalebox{0.8}{
$\begin{array}{l}
\mathbb{P}\left[ {SINR > T|R_{ook}^{} = x} \right]\\
 = \mathbb{P}\left( {1 > \frac{T}{M}(1 + {x^{\alpha \left( {1 - \varepsilon } \right)}}{C_1} + {x^{2\alpha \left( {1 - \varepsilon } \right)}}{C_2} + {C_4}\sum\limits_{j \ne k}^K {{{\left( {{\beta _{00j}}} \right)}^{1 - \varepsilon }}} )} \right)\\
\mathop  \approx \limits^{(a)} \mathbb{P}\left( {g > \frac{T}{M}(1 + {x^{\alpha \left( {1 - \varepsilon } \right)}}{C_1} + {x^{2\alpha \left( {1 - \varepsilon } \right)}}{C_2} + {C_4}\sum\limits_{j \ne k}^K {{{\left( {{\beta _{00j}}} \right)}^{1 - \varepsilon }}} )} \right)\\
\mathop  \approx \limits^{(b)} 1 - E\left\{ {\left( {1 - {e^{ - \frac{{\eta T}}{M}{{\left( {1 + {x^{\alpha \left( {1 - \varepsilon } \right)}}{C_1} + {x^{2\alpha \left( {1 - \varepsilon } \right)}}{C_2} + {C_4}\sum\limits_{j \ne k}^K {{{\left( {{\beta _{00j}}} \right)}^{1 - \varepsilon }}} } \right)}^N}}}} \right)} \right\}\\
\mathop \approx \limits^{(c)} \sum\limits_{n = 1}^N {\left(\!\!\! {\begin{array}{*{20}{c}}
N\\
n
\end{array}} \!\!\! \right){{\left( { - 1} \right)}^{n + 1}}{e^{ - \eta Tn\frac{1}{M}}}Z\left( x \right)L\left( x \right)} \\
\mathop  \Rightarrow \limits^{(d)} \mathbb{P}\left[{SINR > T}\right] = \\
\sum\limits_{n = 1}^N {\left(\!\!\! {\begin{array}{*{20}{c}}
N\\
n
\end{array}} \!\!\! \right){{\left( { - 1} \right)}^{n + 1}}{e^{ - \eta Tn\frac{1}{M}}}} \int_0^\infty  {Z\left( x \right)f\left( x \right){{\left[ {L\left( x \right)} \right]}^{K - 1}}} dx.
\end{array}$
}
\end{equation}
And the the corresponding parameters are
\begin{equation}\label{Parameters-1}
\scalebox{0.8}{
$\begin{array}{l}
{C_1} = \left( {\frac{1}{\Delta } + K} \right){P_1}{I_1} + {\sigma ^2}\\
{C_2} = \left( {K{P_1}{I_1} + {\sigma ^2}} \right)\frac{{{P_1}{I_1}}}{\Delta } + \frac{{\left( {M + 1} \right){P_2}{I_2}}}{\Delta }\\
{C_4} = {x^{\alpha \left( {1 - \varepsilon } \right)}}\left( {1 + {x^{\alpha \left( {1 - \varepsilon } \right)}}\sum\limits_{i \in {\Psi _\lambda }}^{} {\alpha _{ik}^2{{\left( {{\beta _{iik}}} \right)}^{ - \varepsilon }}{{\left( {{\beta _{0ik}}} \right)}^1}} } \right),
\end{array}$
}
\end{equation}
where in ${(a)}$, we use a \emph{dummy} gamma variable $g$ with unit mean and shape parameter $N$ to approximate the constant number one, whose detail is showed in \cite{Tianyang-22};
In ${(b)}$, the approximation follows from Alzer's inequality \cite{Tianyang};
In ${(c)}$, we get the expectation of ${\alpha _{ik}^2}$ and ${\sum\limits_{j\ne k}^K {{{\left({{\beta _{00j}}} \right)}^{1-\varepsilon}}}}$, and ${Z\left(x\right)}$ and ${L\left(x\right)}$ are given in Theorem 1.
Finally ${(d)}$ is because $x$ follows the Rayleigh distribution.


\end{document}